\documentclass{PoS}
\usepackage{amsmath}

\title{Testing reweighting method for truncated Overlap fermions}

\ShortTitle{Testing reweighting method for truncated Overlap fermions}

\author{\speaker{Ken-Ichi Ishikawa}\\
        Graduate School of Science, Hiroshima University,
        Higashi-Hiroshima, Hiroshima 739-8526, Japan\\
        E-mail: \email{ishikawa@theo.phys.sci.hiroshima-u.ac.jp}}

\abstract{
The lattice QCD simulation with the lattice chiral symmetry is very attractive,
however, it is difficult to maintain the symmetry at a modest numerical computation cost.
A candidate to reduce the computational cost during the configuration generation with
the HMC algorithm is to relax the requirement of the chiral symmetry and to use the reweighing 
method recovering the symmetry at the measurement phase.
In this talk, we presented the reweighing method to restore the chiral symmetry
of the truncated overlap fermion operator.
In order to avoid the large discrepancy between the truncated overlap operator and 
the exact overlap operator, we split the reweighting factor into several steps gradually 
increasing the order of truncation. 
We investigated the truncation dependence of the reweighting factor 
on a set of quenched $8^3\times 32$ lattice configurations generated with the DBW2 gauge action.
We found that a large fluctuation on the reweighting factor between 
a high-order truncated overlap operator and the exact overlap operator 
on a couple of configurations.
The origin of the large fluctuation seems to be due to a small 
eigenvalue of the overlap kernel on these configurations.
}

\FullConference{31st International Symposium on Lattice Field Theory - LATTICE 2013\\
		July 29 - August 3, 2013\\
		Mainz, Germany}

\begin{document}

\section{Introduction}
It is a hard task to maintain the lattice chiral symmetry in lattice simulations.
One possibility to reduce the total computational cost is to relax the requirement of 
the chiral symmetry and to use the reweighing method recovering the symmetry at the 
measurement phase.
The HMC algorithm with the truncated overlap fermion with approximate lattice chiral symmetry
has been proposed by Borici in terms of domain-wall type 
fermions~\cite{Borici:1999zw}\cite{Borici:1999da}\cite{Edwards:2000qv}.
The reweighting factor is the determinant ratio between the truncated and exact overlap 
operators and can be estimated by stochastic noise method.
The success of the reweighting method relies on the small fluctuation 
of the reweighting factor on the configuration ensemble and the overlap of 
the statistical histogram between ensemble with the exact and 
with the truncated overlap operators.

The reweighting method for the overlap or domain-wall fermions has been investigated in 
Refs.~\cite{DeGrand:2008ps}\cite{Ishikawa:2010tq}\cite{Oksuzian:2012rzb}\cite{Liu:2012gm} aiming for 
tuning quark masses or improving chiral symmetry.
In Ref.~\cite{Ishikawa:2010tq}, 
the reweighting method has been investigated to
improve the chiral property of the domain-wall simulations. 
In order to tame the fluctuation from the reweighting factor made of
the ratio of two determinants arising from two operators with different fifth-dimensional size,
a determinant splitting via $n$th-root trick and
action parameter tuning were introduced and investigated. They found that
one obstacle to the efficient reweighting comes from the fluctuation among 
configurations.
The quark mass reweighting in the epsilon regime with dynamical overlap fermions 
has been investigated in Ref.~\cite{DeGrand:2008ps} and the method works well.

In this paper we investigate the reweighting method with the truncated overlap fermion
 instead of the domain-wall fermion.
The reweighting factor can be estimated in a four-dimensional form, by which 
the fluctuation from the different truncation level of the overlap operator is reduced. 
The behavior of the reweighing factor and the chiral symmetry violation
on the truncation level is investigated.
Fukaya \textit{et al.} also reported the reweighting technique recovering the chiral 
symmetry with a similar setup~\cite{FUKAYA:LAT2013}.

In the next section we briefly describe 
the truncated overlap operator constructed from the domain-wall operator. 
We also introduce the multiple step reweighting method using the truncated overlap 
fermion. The last step of the reweighting requires 
the overlap operator exact at a numerical precision.
The exact operator is implemented via the optimal domain-wall operator with the 
low-mode improvement~\cite{Hernandez:2000iw}\cite{Jansen:2003jq}.
The even-odd preconditioning for the low-mode improved optimal domain-wall operator
is implemented.
In section \ref{sec:Results} we show the simulation results on
the reweighting factor measured on a quenched QCD $8^3\times 32$ lattice configuration
generated with the DBW2 gauge action.
We summarize this paper in the last section.

\section{Truncated overlap fermion and reweighting method}
\label{sec:Two}
The truncated overlap operator can be realized by several methods depending on
how to approximate the signum function contained in the overlap operator;
(i) Cayley transformation type approximation, (ii) Partial fractional form of 
the rational approximation, (iii) Continued fraction form of the rational approximation~\cite{Edwards:2005an}.
The truncation of the approximation induces a four-dimensional approximated overlap operator (truncated overlap operator).
Any of these approximations can be transformed into the form involving five-dimensional operators.
We employ (i) Cayley transformation approximation which induces the domain-wall type 
five dimensional operator~\cite{Borici:1999zw}\cite{Borici:1999da}\cite{Edwards:2000qv}\cite{Kikukawa:1999sy}.

We employ the following truncated overlap operator;
\begin{equation}
    D_{\mathrm{TROV}}^{N_5} \equiv (P \epsilon)^{\dag}
\left(D_{\mathrm{DWF}}^{(N_5,m_f=1)}\right)^{-1}
D_{\mathrm{DWF}}^{(N_5,m_f)} P \epsilon,
\label{eq:TROV}
\end{equation}
where $P\epsilon$ ($(P\epsilon)^{\dag}$)is the prolongation (restriction) operator 
between five-dimensional and four-dimensional lattices
defined by
\begin{equation}
    P\epsilon=\left( P_L, 0, 0, \cdots, 0,0, P_R\right)^T,
\end{equation}
with $P_L$ and $P_R$ the chiral projection matrices. $N_5$ is the size of the fifth dimension.
$D_{\mathrm{DWF}}^{(N_5,m_f)}$ is the domain-wall operator;
\begin{align}
 D_{\mathrm{DWF}}^{(N_5,m_f)} = D_{\mathrm{WD}} X + Y,&\\
X =
\left(
    \begin{array}{cccccc}
           b_1     &  c_1 P_L & 0       &      0 & \cdots     & -m_f c_1 P_R \\
           c_2 P_R &  b_2     & c_2 P_L &      0 & \cdots     &  0 \\
           0       &  c_3 P_R & b_3     & \ddots & \ddots     &  0 \\
          \vdots   &  \ddots  & \ddots  & \ddots & \ddots     & \vdots \\
           0       &  0       & 0       & \ddots & b_{N_5-1}  & c_{N_5-1} P_L \\
  -m_f c_{N_5} P_L &  0       & 0       & \cdots & c_{N_5}P_R & b_{N_5} 
    \end{array}
\right),& \quad\quad Y = \left.X\right|_{b_j=1,c_j=-1},
\label{eq:DWFOP}
\end{align}
where $X$ has the indexes in the spin and fifth-direction.
$D_{\mathrm{WD}} $ is the standard four-dimensional Wilson-Dirac operator with 
a negative mass $M$, and will be multiplied on each element of $X$. 
We omit the identity matrix
with the color, spin, and four-dimensional site indexes in front of $Y$.
The parameters $b_j$ and $c_j$ are tunable parameters introduced
for the M\"{o}bius domain-wall operator in Ref.~\cite{Brower:2005qw}.

The M\"{o}bius domain-wall operator interpolates between 
Shamir's standard domain-wall operator and Bori\c{c}i's domain-wall 
implementation of Neuberger's overlap operator 
including Chiu's optimal operators~\cite{Chiu:2002ir}. 
We employ the following choices:
\begin{description}
\item[Type-A] Shamir's standard domain-wall operator: $b_j = a_5$, $c_j=0$.
\item[Type-B] Optimal Shamir's standard domain-wall operator: $b_j = (w_j+a_5)/2$, $c_j=(w_j-a_5)/2$.
\end{description}
where $a_5$ is a tunable normalization constant and $w_j$'s are the coefficients 
from the Zolotarev optimal approximation for the signum function~\cite{Chiu:2002ir}. 
These choices realize the following overlap operator in the infinite $N_5$ limit.
\begin{eqnarray}
D_{\mathrm{OV}}= \lim_{N_5\to\infty} D^{N_5}_{\mathrm{TROV}}  &= &
\frac{1+m_f}{2}+\frac{1-m_f}{2} \gamma_5 \mathrm{sign}({\cal H}_{W}),\\
{\cal H}_{W} &=& \gamma_5 D_{\mathrm{WD}} \left(a_5 D_{\mathrm{WD}} / 2 + 1\right)^{-1}.
\label{eq:Kernel}
\end{eqnarray}
The \textbf{Type-A} coefficient is used for the truncated overlap operator $D^{N_f}_{\mathrm{TROV}}$,
while the \textbf{Type-B} coefficient is for the exact overlap operator $D_{\mathrm{OV}}$.

Since the configuration generation with the exact overlap operator is a hard task,
we employ the truncated overlap operator with a modest $N_5$ in the configuration generation. 
The domain-wall fermion simulations with a modest $N_5$ seem to realize 
a small residual masses compared to the QCD scale and lightest quark masses 
at the vanishing pseudo-scalar meson mass~\cite{Allton:2007hx}.
Thus we expect that the ensemble generated with the truncated overlap operator 
with a modest $N_5$ could have sufficient statistical overlap to the ensemble with 
the exact overlap operator.
The reweighting method enables us to evaluate the expectation value with the exact chiral symmetry
using the ensemble generated with the truncated overlap operator. 
The expectation value of an observable $O$ in the two-flavor QCD case can be evaluated as
\begin{align}
    \langle O \rangle =&
\left\langle O \prod_{j=1}^{N_{\mathit{step}}} 
W_{(j)} 
\right\rangle_{\mathrm{TROV}(N_{\mathrm{step}})}
\bigg/
\left\langle \prod_{j=1}^{N_{\mathit{step}}}
W_{(j)}
\right\rangle_{\mathrm{TROV}(N_{\mathrm{step}})}, \\
W_{(1)} &= \det\left[ D_{\mathrm{OV}} / D_{\mathrm{TROV}}^{N_{5(1)}}\right]^2,
\label{eq:REWFAC1}
\\
W_{(j)} &= \det\left[ D_{\mathrm{TROV}}^{N_{5(j-1)}} / D_{\mathrm{TROV}}^{N_{5(j)}}\right]^2 
\quad\quad \mbox{(for $j = 2,3,\cdots, N_{\mathrm{step}}$)}.
\label{eq:REWFAC2}
\end{align}
The sequence of the size of fifth dimension $N_{5(j)}$ is chosen to satisfy
\begin{equation}
    N_{5(1)} >  N_{5(2)} > N_{5(3)} > \cdots > N_{5(N_{\mathrm{step}}-1)} > N_{5(N_{\mathrm{step}})}.
\end{equation}
The reweighting factors, $W_{(j)}$ ($j=2,3,\cdots,N_{\mathrm{step}}$), are 
evaluated with the \textbf{Type-A} coefficient.
The expectation values with $
\left\langle 
\cdots
\right\rangle_{\mathrm{TROV}(N_{\mathrm{step}})}$ are evaluated 
on the configuration ensemble generated with the truncated overlap operator
$D_{\mathrm{TROV}}^{N_{5(N_\mathrm{step})}}$. 
The determinants are evaluated with the stochastic estimator with the Gaussian random vector $\chi$;
\begin{equation}
    W_{(j)}=\langle \exp\left(-dS\right) \rangle_{\chi},
\quad\quad
dS \equiv \left|
 \left(D_{\mathrm{TROV}}^{N_{5(j-1)}}\right)^{-1} D_{\mathrm{TROV}}^{N_{5(j)}}\chi\right|^2 - \left|\chi\right|^2.
\label{eq:NoiseEstimate}
\end{equation}
We expect that the determinant is close to 1 with a small interval step.
The determinant splitting reduces the fluctuation from the stochastic estimator 
on each configuration~\cite{Hasenfratz:2008fg}\cite{Ishikawa:2010tq}.
The success of the reweighting method depends on the size of fluctuation of each $W_{(j)}$ 
among configuration.

In the last step of the determinant splitting we need the exact overlap operator $D_{\mathrm{OV}}$
in $W_{(1)}$. To ensure that it holds the exact lattice chiral symmetry at double precision,
we employ the optimal \textbf{Type-B} coefficient combined with the low-mode 
improvement method~\cite{Hernandez:2000iw}\cite{Jansen:2003jq}.
After computing several low-modes and high-modes of the kernel operator ${\cal H}_{W}$ Eq.~(\ref{eq:Kernel}),
we shift the low-modes according to the prescription of Ref.~\cite{Jansen:2003jq}, and 
we determine the Zolotarev coefficients $w_j$'s and the size of fifth dimension $N_5$ 
optimal for the shifted eigenvalue range of the kernel operator. 
The low-mode improved optimal domain-wall operator 
$\tilde{D}^{imp}_{\mathrm{DWF}}$ can be written as
\begin{align}
  &\tilde{D}^{imp}_{\mathrm{DWF}} \equiv D^{imp}_{\mathrm{DWF}}X^{-1}
   = \tilde{D}_{\mathrm{DWF}} + W_k S_k^{\dag},
\label{eq:ImpDWF}\\
  & \tilde{D}_{\mathrm{DWF}} \equiv D_{\mathrm{DWF}}X^{-1} = K - \frac{1}{2} M_{\mathrm{hop}},
  & K= (4-M)+Y X^{-1},\\
  &{\cal H}_{\mathrm{W}} V_k = V_k \Lambda_k, 
  &  \Lambda_k = \mathrm{diag}(\lambda_1,\lambda_2,\cdots,\lambda_k),\\
  &  W_k= (a_5D_{\mathrm{WD}}/2+1) \gamma_5 V_k,
  &  S_k = \mathrm{diag}(\hat{\lambda}_1,\hat{\lambda}_2,\cdots, \hat{\lambda}_k)-\Lambda_k,
\end{align}
where we multiply $X^{-1}$ to explicitly extract the four-dimensional 
hopping matrix $M_{\mathrm{hop}}$.
In this form the operator can be regarded as the $N_5$-flavor four-dimensional operator 
with the spin-flavor mixing mass matrix $K$.
$V_k$ spans the $k$-dimensional low-mode eigen space of 
the kernel operator ${\cal H}_W$. $\hat{\lambda}_j$ are shifted eigenvalues 
$\hat{\lambda}_j=2\mathrm{sign}(\lambda_j)\max_{s=1,\cdots,k}|\lambda_s|$.
The four-dimensional even/odd site preconditioning can be applied not only to
$\mathrm{\tilde{D}_{\mathrm{DWF}}}$ but also to the improved operator Eq.~(\ref{eq:ImpDWF}) 
with the type-B coefficient. 
In order to apply the even/odd site preconditioning to 
the linear equation 
\begin{equation}
\tilde{D}^{imp}_{\mathrm{DWF}} x = b,
\end{equation}
we introduce an auxiliary vector $\zeta_k$ with $k$ componetns,
and reorder the unknowns of the linear equation to yield the following blocked form;
\begin{equation}
\left(
    \begin{matrix}
(\tilde{D}_{\mathrm{DWF}})_{ee} &
(\tilde{D}_{\mathrm{DWF}})_{eo} &
(W_k)_e\\
(\tilde{D}_{\mathrm{DWF}})_{oe} &
(\tilde{D}_{\mathrm{DWF}})_{oo} &
(W_k)_o\\
(S^{\dag}_k)_e &
(S^{\dag}_k)_o &
-1 
    \end{matrix}
\right)
\left(
    \begin{matrix}
        x_e \\ x_o \\ \zeta_k
    \end{matrix}
\right)=
\left(
    \begin{matrix}
        b_e \\ b_o \\ 0
    \end{matrix}
\right),
\end{equation}
where suffixes $e$ and $o$ mean the lattice site even/odd-ness.
By eliminating $\zeta_k$ and 
$x_o$ from the blocked equation, we can obtain the even/odd site 
preconditioned linear equation for $x_e$.

\section{Results}
\label{sec:Results}

We investigate the behavior of the reweighing factor
Eqs~(\ref{eq:REWFAC1}) and (\ref{eq:REWFAC2}) 
and the chiral symmetry violation due to the truncation of the overlap operator.
We generate the quenched $SU(3)$ gauge configurations using the DBW2 gauge action 
with $\beta=0.87$ on a $8^3\times 32$ lattice. 
We use the 10 configurations separated by HMC 100 trajectories.
We scan $N_5$ from 12 to 32 with the constant step size $N_{5(j-1)}-N_{5(j)}=2$.
We employ $(m_f=0.02, M=1.8)$ for the domain-wall fermion parameter.
We observe $M_{PS}=0.274(21)$ and $M_{V}=0.661(65)$ at $N_5=12$ for the pseudo-scalar 
and vector meson masses respectively, which correspoinds 
to $a^{-1}\sim 1.3$ GeV~\cite{Aoki:2002vt}.

The twelve low and high eigen modes of the kernel ${\cal H}_W$ (\ref{eq:Kernel}) are measured
to construct $D_{\mathrm{OV}}$ as in Fig.~\ref{fig:EigHW}. 
Isolated small eigen values appear in some configurations.
Such small eigen values affect the quality of the chiral symmetry of the truncated operator.
In order to see the chiral symmetry violation we measure the following metric;
\begin{equation}
\Delta_{\mathrm{GW}}=
\eta^{\dag}\left[
  \{\gamma_5, D^{-1} \}-\frac{2}{1+m_f}\left( m_f D^{-1}\gamma_5 D^{-1} + \gamma_5\right)
\right]\eta,
    \label{eq:DeltaGWR}
\end{equation}
where $\eta$ is the Gaussian random noise vector with unit norm and 
$D=D_{\mathrm{OV}}$ or $D=D_{\mathrm{TROV}}^{N_5}$.
Figure~\ref{fig:DeltaGW} shows the chiral symmetry violation $\Delta_{\mathrm{GW}}$ measured 
with four noises on each configuration.
The accuracy of the Zolotarev approximation for the signum function we used
is at $10^{-13}$ and the optimal operator achieves $10^{-12}$ accuracy.
The $N_5$ dependence of $\Delta_{\mathrm{GW}}$ on the configurations 
having small eigen values in ${\cal H}_W$ is small, 
which indicates the slow convergence to the exact overlap operator 
on such configurations.

Figs.~\ref{fig:expdSNs12} and \ref{fig:expdSNs30} show the 
exponential weight $\exp(-dS)$ in Eq.~(\ref{eq:NoiseEstimate}) 
before taking noise average. Two noises are generated on each configuration.
The factor between the truncated overlap operators at low order ($N_5=12$ and $N_5=14$) 
has large values (black circles in Fig.~\ref{fig:expdSNs12})
while  at high order  ($N_5=30$ and $N_5=32$) it has
$O(1)$ values  (black circles in Fig.~\ref{fig:expdSNs30}). 
The fluctuation between truncated operators (black circles) is 
mild both on the noise dependence and on the configuration dependence.

As seen from the red triangles in the figures, however, 
the fluctuation of the factor between the truncated operator and the exact 
operator is very large. We note that the large fluctuation occurs on 
the configurations which contain small eigen values in 
${\cal H}_W$ (config.\#= 1, 2, 3, 9, and 10).
This large fluctuation could spoil the validity of the use of 
the reweighting method for recovering the exact chiral symmetry.

\newcommand{\figscale}{2.95in}

\begin{figure}[t]
 \begin{minipage}{0.48\hsize}
      \centering
\includegraphics[width=\figscale]{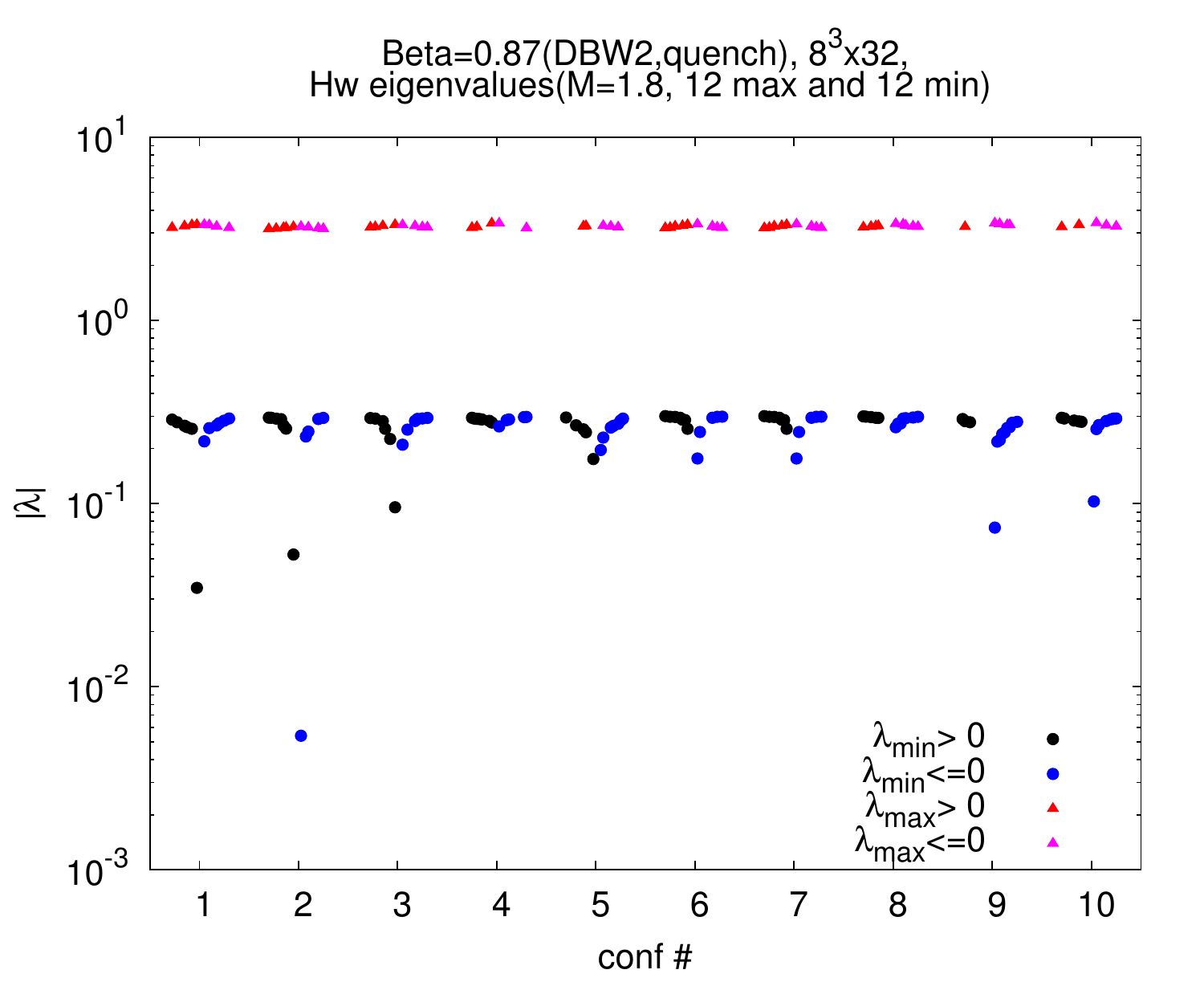}
\caption{Twelve high and low eigen values of the kernel operator ${\cal H}_W$ on each configuration.}
\label{fig:EigHW}
 \end{minipage}
\hfil
 \begin{minipage}{0.48\hsize}
      \centering
\includegraphics[width=\figscale]{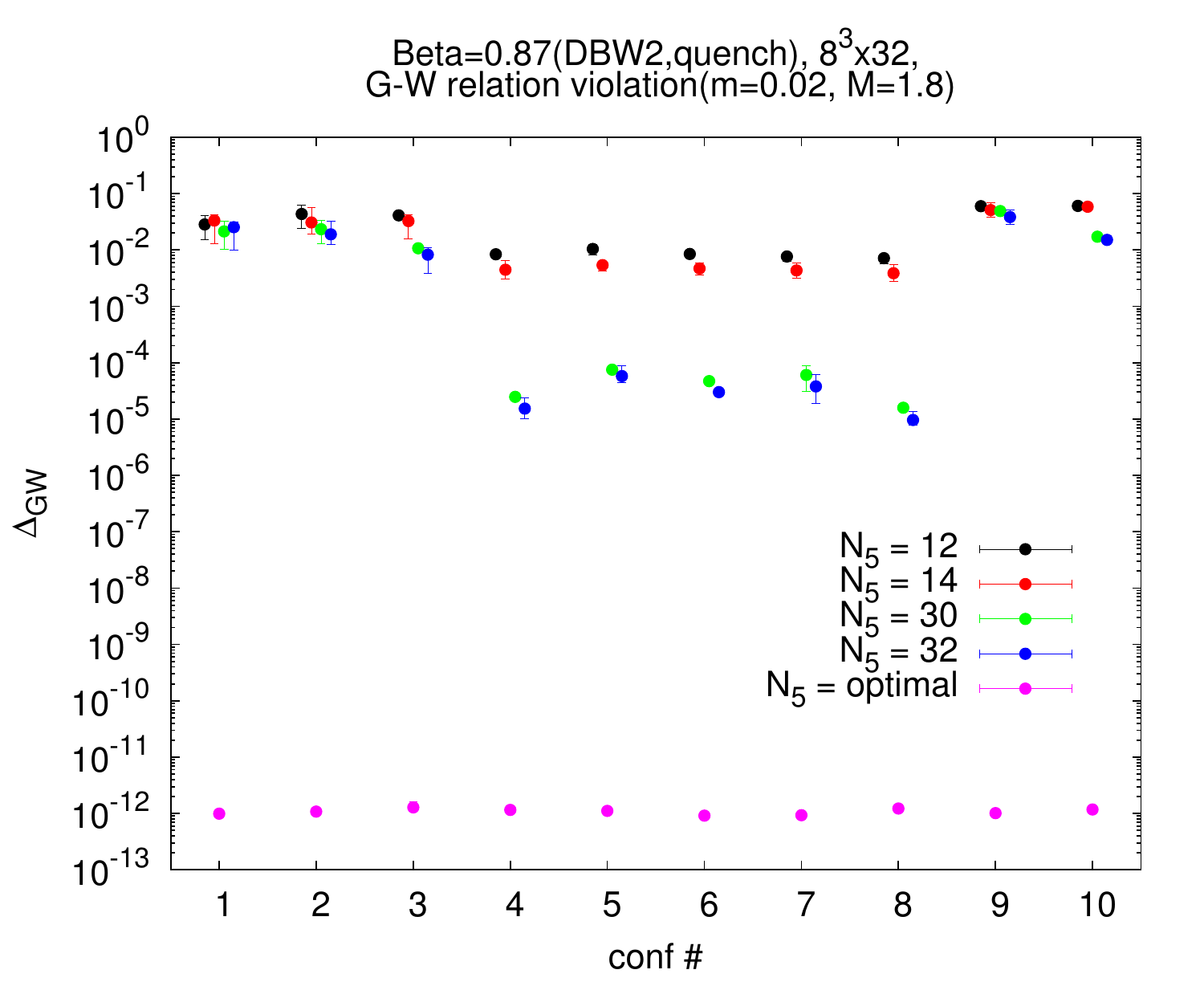}
\caption{Violation of the Ginsparg-Wilson relation $\Delta_{\mathrm{GW}}$ on each configuration.}
\label{fig:DeltaGW}
 \end{minipage}
\end{figure}

\begin{figure}[t]
 \begin{minipage}{0.48\hsize}
      \centering
\includegraphics[width=\figscale]{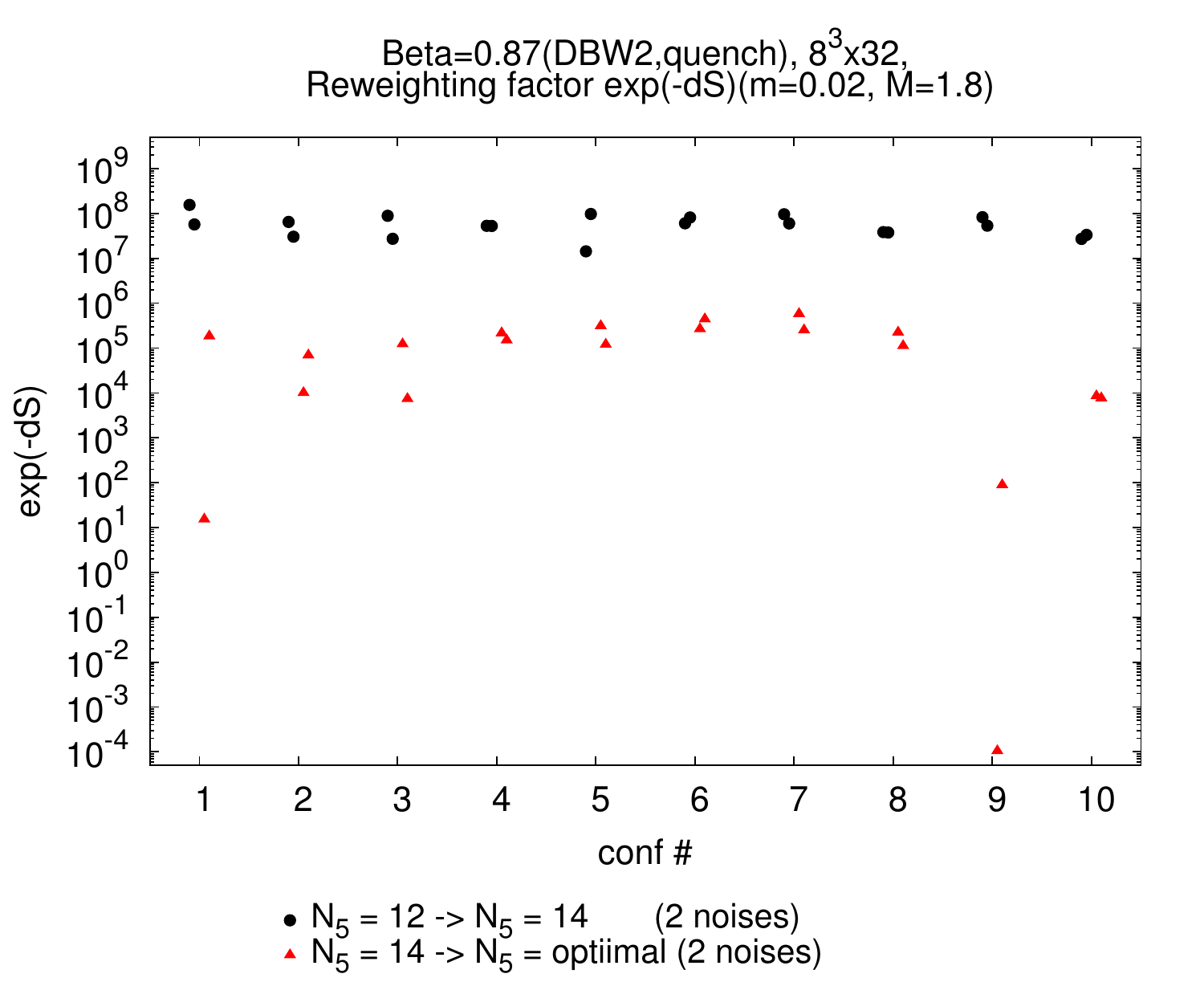}
\caption{Reweighting factor $\exp(-dS)$ before noise averaging on each configuration.}
\label{fig:expdSNs12}
 \end{minipage}
\hfil
 \begin{minipage}{0.48\hsize}
      \centering
\includegraphics[width=\figscale]{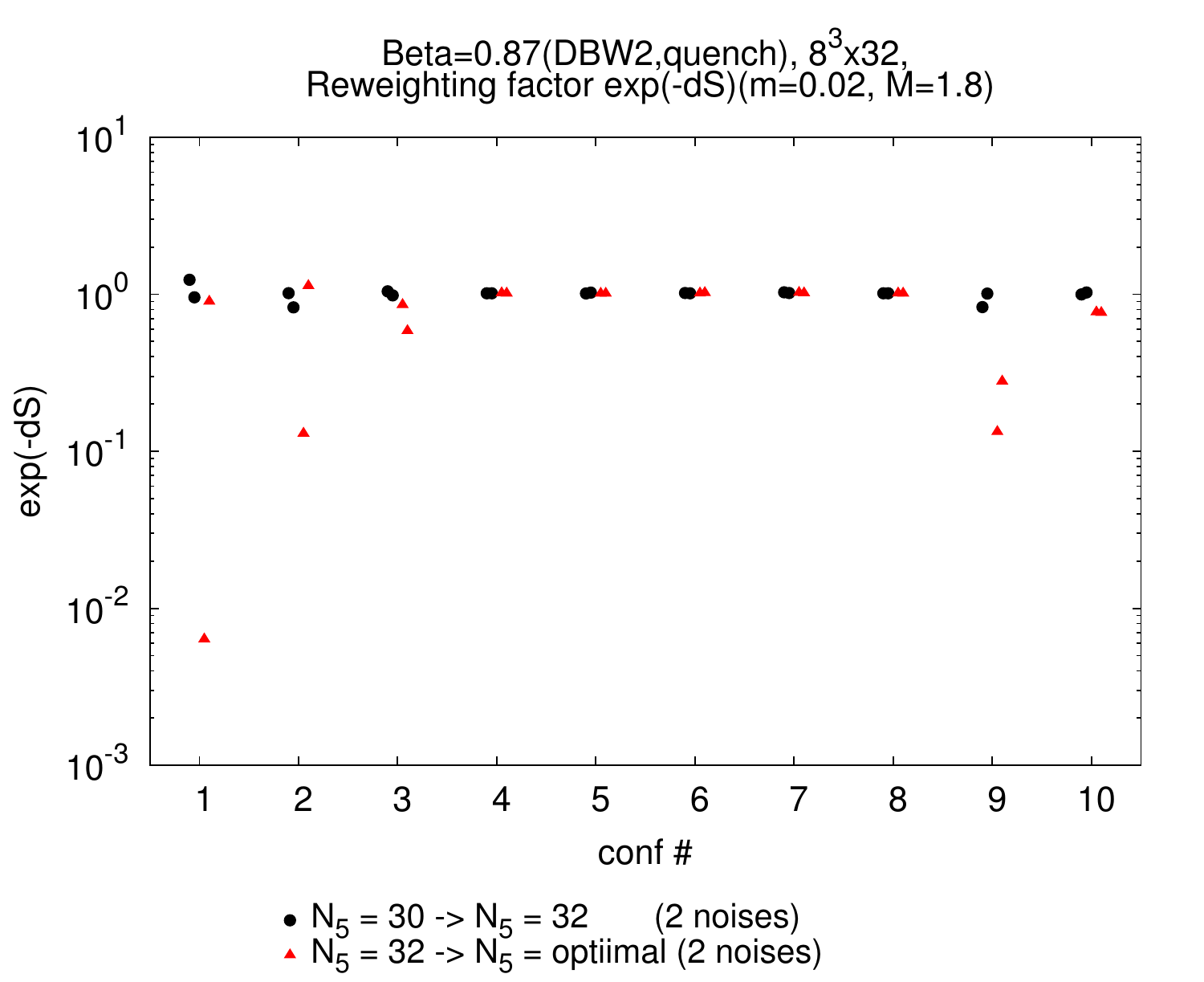}
\caption{Reweighting factor $\exp(-dS)$ before noise averaging on each configuration.}
\label{fig:expdSNs30}
 \end{minipage}
\end{figure}

\section{Summary}
\label{sec:Summary}
We have investigated 
the reweighting method aiming for restoring the chiral symmetry of the truncated
overlap operator. Using a set of quenched configurations, we have observed
a clear correlation among 
the appearance of small eigen values in the overlap kernel operator
${\cal H}_W$, the chiral symmetry violation, and the fluctuation of 
the reweighting factor between the truncated and exact overlap operators.
The appearance of the small eigen modes or near zero modes 
in the kernel operator reflects the change of the topological property of 
the gauge configuration and zero modes of the mass-less exact overlap operator.
A more detailed analysis on the relation between the topological zero modes and 
the approximation of the overlap operator has been carried out by 
Fukaya \textit{et al.}\cite{FUKAYA:LAT2013}, where the mismatch between
the zero mode of the overlap operator and the truncated overlap operator has been reported.

It seems that the problems arising from the near zero modes of 
the kernel operator ${\cal H}_W$ still remain even if we employ 
the reweighting method.

\paragraph*{Acknowledgment}
The numerical computations have been performed using a PC-cluster of HPCI 
(High Performance Computing Infrastructure)
strategic program Field 5 at the Center for Computational Sciences, University of Tsukuba,
and a PC-cluster of INSAM (Institute for Numerical Simulations and Applied Mathematics) 
at Hiroshima University.
This work was supported in part by a Grant-in-Aid for Scientific Research (C)
(No. 24540276) from the Japan Society for the Promotion of Science (JSPS).


\begin{thebibliography}{10}

\bibitem{Borici:1999zw}
  A.~Borici,
  Nucl.\ Phys.\ Proc.\ Suppl.\  {\bf 83} (2000) 771
  [hep-lat/9909057].

\bibitem{Borici:1999da}
  A.~Borici,
  hep-lat/9912040.

\bibitem{Edwards:2000qv}
  R.~G.~Edwards and U.~M.~Heller,
  Phys.\ Rev.\ D {\bf 63} (2001) 094505
  [hep-lat/0005002].

\bibitem{DeGrand:2008ps}
  T.~DeGrand,
  Phys.\ Rev.\ D {\bf 78} (2008) 117504
  [arXiv:0810.0676 [hep-lat]].

\bibitem{Ishikawa:2010tq}
  T.~Ishikawa, Y.~Aoki and T.~Izubuchi,
  PoS LAT {\bf 2009} (2009) 035
  [arXiv:1003.2182 [hep-lat]].

\bibitem{Oksuzian:2012rzb}
  H.~Ohki {\it et al.}  [JLQCD Collaboration],
  Phys.\ Rev.\ D {\bf 87} (2013) 3,  034509
  [arXiv:1208.4185 [hep-lat]].

\bibitem{Liu:2012gm}
  Q.~Liu, N.~H.~Christ and C.~Jung,
  Phys.\ Rev.\ D {\bf 87} (2013) 5,  054503
  [arXiv:1206.0080 [hep-lat]].

\bibitem{FUKAYA:LAT2013}
  H.~Fukaya, S.~Aoki, G.~Cossu, S.~Hashimoto, T.~Kaneko and J.~Noaki,
  in these proceedings.

\bibitem{Hernandez:2000iw}
  P.~Hernandez, K.~Jansen and M.~Luscher,
  hep-lat/0007015.

\bibitem{Jansen:2003jq}
  K.~Jansen and K.~-i.~Nagai,
  JHEP {\bf 0312} (2003) 038
  [hep-lat/0305009].

\bibitem{Edwards:2005an}
  R.~G.~Edwards, B.~Joo, A.~D.~Kennedy, K.~Orginos and U.~Wenger,
  PoS LAT {\bf 2005} (2006) 146
  [hep-lat/0510086].

\bibitem{Kikukawa:1999sy}
  Y.~Kikukawa and T.~Noguchi,
  hep-lat/9902022.

\bibitem{Brower:2005qw}
  R.~C.~Brower, H.~Neff and K.~Orginos,
  Nucl.\ Phys.\ Proc.\ Suppl.\  {\bf 153} (2006) 191
  [hep-lat/0511031].

\bibitem{Chiu:2002ir}
  T.~-W.~Chiu,
  Phys.\ Rev.\ Lett.\  {\bf 90} (2003) 071601
  [hep-lat/0209153].

\bibitem{Allton:2007hx}
  C.~Allton {\it et al.}  [RBC and UKQCD Collaborations],
  Phys.\ Rev.\ D {\bf 76} (2007) 014504
  [hep-lat/0701013].

\bibitem{Hasenfratz:2008fg}
  A.~Hasenfratz, R.~Hoffmann and S.~Schaefer,
  Phys.\ Rev.\ D {\bf 78} (2008) 014515
  [arXiv:0805.2369 [hep-lat]].

\bibitem{Aoki:2002vt}
  Y.~Aoki, T.~Blum, N.~Christ, C.~Cristian, C.~Dawson, T.~Izubuchi, G.~Liu and R.~Mawhinney {\it et al.},
  Phys.\ Rev.\ D {\bf 69} (2004) 074504
  [hep-lat/0211023].

\end{thebibliography}
\end{document}